\documentclass[11pt]{article}
\usepackage{graphicx} 
\usepackage{utopia}
\usepackage{fullpage}
\usepackage{xspace}
\usepackage{url}
\usepackage{color}
\newcommand{\lfshort}[0]{\textsc{LF}\xspace}

\newcommand{\lf}[0]{\textsc{Lingua Franca}\xspace}

\definecolor{fixmeColor}{rgb}{0.8,0.1,0.1}

\definecolor{stringColor}{rgb}{0.8,0.1,0.1}
\usepackage{listings}
\usepackage{courier}
\definecolor{airforceblue}{rgb}{0.36, 0.54, 0.66}
\definecolor{bluegray}{rgb}{0.4, 0.6, 0.8}
\definecolor{darkcerulean}{rgb}{0.03, 0.27, 0.49}
\lstdefinelanguage{LF}{
  keywords={deadline, after, state, logical, physical, startup, shutdown, reaction, preamble, target, reactor, trigger, input, output, constructor, new, action, clock, actor, handler, time, main, federated, timer, sec, secs, msec, msecs, usec, usecs, mode, initial, implements, import, if, while},
  emph={L,name, type, init, effect, instance}, emphstyle=\itshape,
  keywordstyle=\color{black}\bfseries,
  ndkeywords={lf_set, lf_set_mode},
  ndkeywordstyle=\color{darkcerulean}\bfseries,
  identifierstyle=\color{black},
  sensitive=false,
  comment=[l]{//},
  morecomment=[s]{/*}{*/},
  commentstyle=\color{airforceblue}\ttfamily,
  stringstyle=\color{black}\ttfamily,
  morestring=[b]',
  morestring=[b]"
}
\title{Exploration of Approaches for Robustness and Safety in a Low~Code Open Environment for Factory Automation -- Technical~Report}

\author{
    Gustavo Quiros A., Yi Peng Zhu, Tao Cui
    \\ Siemens Technology \\ 
    \{gustavo.quiros,yipeng.zhu,tao.cui\}@siemens.com 
    \and 
    Shaokai Lin, Marten Lohstroh, Edward Lee    
    \\ University of California, Berkeley \\ 
    \{shaokai,marten,eal\}@berkeley.edu 
    }

\date{September 2024}

\begin{document}

\maketitle

\begin{abstract}
This report is a compilation of technical knowledge and concepts that were produced by the authors and additional contributors in the context of the collaboration projects "Abstraction Requirements for Language of Choice in Industrial Automation" (FY21-22) and "Approaches for Robust and Safe Low-Code" (FY23-24) from Siemens Technology and the University of California, Berkeley. The primary objective of these projects was to assess Siemens Open Industrial Edge (OIE) engineering capabilities by defining a concept that ensures the satisfaction of coordination and safety requirements when using disparate OIE modules. The objective was to use the Lingua Franca (LF) coordination language to demonstrate how to address challenges in: 1. engineering modular, distributed, and flexible automation solutions that ensure, by design, robust and safe operation1; 2. the use of IEC 61499, the event driven execution model for specifying the execution order of OIE modules (defined as function blocks); 3. support large-scale distributed OIE automation solutions, and eventually 4. define optimal solutions with synchronization and time-optimal mechanisms.
\end{abstract}

\section{Introduction} 

Digitalization is disrupting business models with new value propositions to reduce costs, improve customer experience, and increase profitability. Open automation promises to accelerate the adoption of digitalization across the value chain by identifying more intuitive and reliable solutions to the costly problem of technological lock-in; a condition that results when unique proprietary solutions are propagated. For this purpose, Siemens launched the Open Industrial Edge Ecosystem (OIE) [1], a digital and vendor independent, cross-manufacturer platform for industry customers. With OIE, customers can benefit from a broad range of software components, offered by numerous providers and manufacturers, that they can integrate into their production processes in a standardized manner. Example applications range from connectivity, data storage, visualization, and analysis right up to machine monitoring, as well as energy and asset management.  

From an engineering standpoint, based on a given manufacturing purpose, the automation engineer brings different modules from the OIE platform and stitches them together to generate an automation solution. However, how to guarantee the correctness and safety, and eventually optimality of the overall behavior, are relevant issues that arise when using disparate distributed modules. Safety modules must be designed with redundancy integrated into them, and verification approaches for safe behavior must be conducted to ensure that an unsafe condition is not introduced. Safety modules must meet specific requirements per ISO-13849 [2] and IEC-62061 [3] standards. These modules must receive specific safety integrity level (SIL) and performance level-e (PL e) ratings to provide traceability and authenticity to their design and build for use in safety systems. On the other hand, optimality has great impact on system performance. In large-scale systems with a multitude of composite modules, it is critical to define optimal solutions with synchronization and time-optimal mechanisms. 

This paper is a compilation of technical knowledge and concepts that were produced by the authors and additional contributors in the context of the collaboration projects ``Abstraction Requirements for Language of Choice in Industrial Automation'' (FY21-22) and ``Approaches for Robust and Safe Low-Code'' (FY23-24) from Siemens Technology and the University of California, Berkeley. The primary objective of these projects was to assess Siemens OIE engineering capabilities by defining a concept that ensures the satisfaction of coordination and safety requirements when using disparate OIE modules. The objective was to use the Lingua Franca (LF) coordination language to demonstrate how to address challenges in:
\begin{enumerate}
    \item engineering modular, distributed, and flexible automation solutions that ensure, by design, robust and safe operation\footnote{Example safety requirements can be found in [4].};
    \item the use of IEC 61499 [4], the event driven execution model for specifying the execution order of OIE modules (defined as function blocks);
    \item support large-scale distributed OIE automation solutions, and eventually
    \item define optimal solutions with synchronization and time-optimal mechanisms. 
\end{enumerate} 

\subsection{Acknowledgements}

The authors would like to thank the following collaborators for providing various contributions reflected the material that is included in this report: Alexander Schulz-Rosengarten, Reinhard von Hanxleden, Soroush Bateni, Benjamin Asch, Ankit Shukla, Yassine Qamsane, Xiao Bo Yang, Jian Qiang Wu, Yue Rong Li, Max Wang, Peter Mertens, Florian Ersch.


\section{Overview of \lf}\label{sec:lf}

\lf (\lfshort)~\cite{LohstrohEtAl:21:Towards} is a polyglot coordination language designed to augment multiple mainstream programming languages (also called target languages), currently C, C++, Python, TypeScript, and Rust, with deterministic reactive concurrency and the capability to specify timed behavior.
\lfshort is supported by a runtime system that enables concurrent and distributed execution of reactive programs, which can be deployed on various platforms, including in the cloud, at the edge, in containers, and even on resource-constrained bare-metal embedded platforms.

A \lf program defines interactions between components known as reactors~\cite{Lohstroh:2019:CyPhy}, with the logic for each reactor written in plain target code.
\lf's code generator then produces one or more programs in the target language, which are compiled using standard tool chains.
When the application has parallelism, it runs on multiple cores without losing determinism.
For distributed applications, multiple programs and scripts are generated to deploy these programs on multiple machines and/or containers.
The network communication fabric connecting these components is also synthesized as part of the code generation and compilation process.

\subsection{Reactor-Oriented Programming}
\lf programs consist of reactors, which are stateful entities with event-driven routines.
Reactors adopt advantageous semantic features from established models of computation, particularly actors~\cite{Agha:97:Actors}, logical execution time~\cite{Kirsch:12:LET}, synchronous reactive languages~\cite{benveniste2003synchronous}, and discrete event systems~\cite{LeeEtAl:14:DE} (such as DEVS~\cite{zeigler1997devs} and SystemC~\cite{liao1997efficient}).
The reaction routines belonging to reactors can process inputs, generate outputs, alter the reactor's state, and schedule future events. Reactors resemble actors~\cite{Agha:97:Actors}, which are software components that communicate through message passing.
However, unlike traditional actors, these messages have timestamps, and the concurrent interaction of reactors is deterministic by default.
Any nondeterministic behavior must be explicitly programmed if needed.

\begin{figure}
	\centering
	\includegraphics[width=\columnwidth]{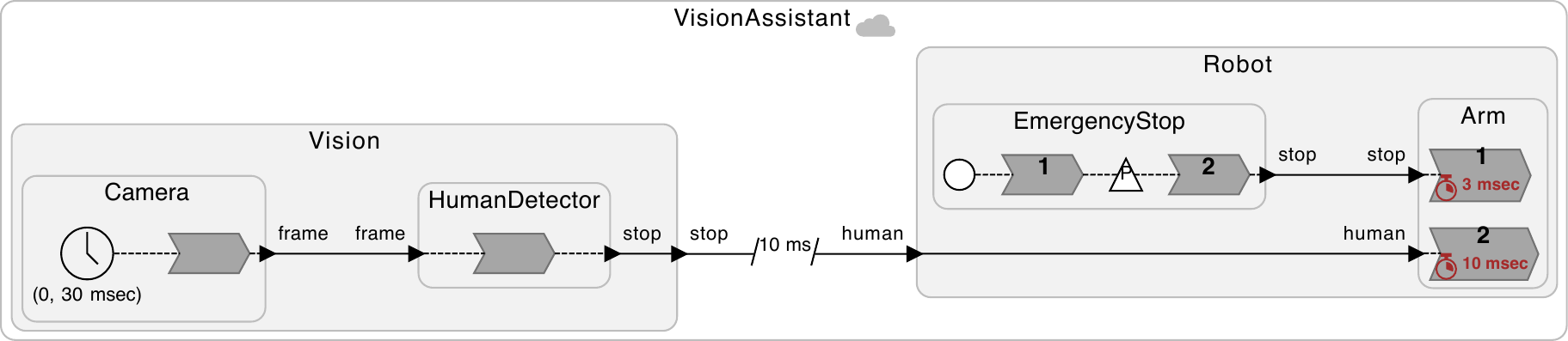}
	\begin{minipage}{0.49\columnwidth}
		\begin{lstlisting}[basicstyle=\ttfamily\scriptsize,language=LF,escapechar=|]
target C // or Python, Rust, etc.|\label{ln:target}|
federated reactor {|\label{ln:federated}|
  robot = new Robot()
  vision = new Vision()
  vision.stop -> robot.human after 10 ms |\label{ln:after}|
}
reactor Robot {
  input human: void
  pedal = new EmergencyStop()
  stop = new Arm()
  pedal.stop -> stop.stop
  human -> stop.human
}
reactor Vision {|\label{ln:vision}|
  output stop: void
  camera = new Camera()
  detect = new HumanDetector()
  camera.frame -> detect.frame
  detect.stop -> stop
}
reactor Camera {
  timer t(0, 30 ms)|\label{ln:timer}|
  output frame: void
  reaction(t) -> frame {=
    // Target language code here...|\label{ln:body1}|
  =}
}
    \end{lstlisting}
  \end{minipage}
  \begin{minipage}{0.49\columnwidth}
    \begin{lstlisting}[basicstyle=\ttfamily\scriptsize,language=LF,escapechar=|,firstnumber=29]
reactor HumanDetector {
  input frame: void
  output stop: void
  reaction(frame) -> stop {=
    // Target language code here...|\label{ln:body2}|
  =}
}
reactor EmergencyStop {
  physical action button |\label{ln:physical}|
  output stop: void
  reaction(startup) -> button {=
    // Target language code here...|\label{ln:startup}|
  =}
  reaction(button) -> stop {=
    // Target language code here...|\label{ln:brake}|
  =}
}
reactor Arm {
  input human: void
  input stop: void
  reaction(stop) {=
    // Target language code here...
  =} deadline(3 ms) {=|\label{ln:pedal_deadline}|
    // Target language code here...|\label{ln:deadline1}|
  =}
  reaction(human) {=
    // Target language code here...
  =} deadline(10 ms) {=|\label{ln:assist_deadline}|
    // Target language code here...|\label{ln:deadline2}|
  =}
}
		\end{lstlisting}
	\end{minipage}
	\caption{Robot system schematic and \lf specification.}
	\label{fig:adas}
\end{figure}

\figurename~\ref{fig:adas} shows a small illustrative example \lf program, \texttt{VisionAssistant}, giving the architecture of
a computer vision system augmenting a robot safety controller.
The job of the vision system is to detect humans and to have the robot react by switching to a safer mode of operation (or shutting down).
This is put in parallel with a more conventional emergency stop subsystem.

The textual code shown at the bottom of the figure is \lf code, and the diagram above it is automatically generated by the tools and updated dynamically as the code is edited.
In this example, the main program is \textit{federated} (line \ref{ln:federated}), which means that each of the top-level reactors, \texttt{Vision} and \texttt{Robot}, will be code generated into its own program in the target language (specified on line \ref{ln:target}).
These programs, called \textit{federates}, can be containerized for better isolation and fault tolerance and/or distributed to distinct hardware, for example to exploit specialized hardware in the \texttt{Vision} reactor.
This system includes a camera with a computer vision system that detects humans and notifies the robot controller when one is identified.

The \texttt{Vision} federate, defined starting on line \ref{ln:vision}, contains two child reactors: \texttt{Camera}, whose reaction is triggered by a timer to periodically capture images, and \texttt{HumanDetector}, which performs a vision task for detecting humans in the captured images.
The grey chevrons in inside reactors in the diagram represent \textit{reactions}, which are triggered by inputs, timers, or events on an event queue and are able to produce outputs.
The business logic inside reactions is written in the target language that is chosen.
A typical implementation of these reactors could use, for example, the Python target, which can leverage existing packages such as Tensorflow lite for the vision component, if it is to be realized on the edge, or it could use an API to realize the vision component in the cloud, or it could use the C target together with OpenCV, the open computer vision library.
Legacy code and libraries are easy to use in \lfshort by simply invoking their APIs in the reaction bodies on lines \ref{ln:body1} and \ref{ln:body2}.

In principle, a federated \lfshort program could use different target languages for each of the federates, such as Python for the \texttt{Vision} federate and C or Rust for the \texttt{Robot} component, but this capability only exists currently in concept demonstration form.

\texttt{Robot} includes two child reactors, \texttt{EmergencyStop} for capturing the asynchronous events triggered by pushing an emergency stop button and \texttt{Arm} for taking inputs from the human detector and stop input so as to control the robot arm.
The  \texttt{EmergencyStop} reactor has a \textit{physical action}, defined on line \ref{ln:physical}, which is used to inject asynchronous external events from the environment.
The reaction to the \textit{startup} event on line \ref{ln:startup} can be used to set up any external interactions, for example enabling an interrupt service routine to handle emergency stop events and schedule the physical action.
The code on line \ref{ln:brake} will then be invoked in reaction to those events.

The \texttt{Arm} reactor includes two \textit{deadline} declarations, which serve two purposes.
First, they guide the \lfshort scheduler to prioritize reaction invocations with nearer deadlines.
Second, they provide fault handling code, on lines \ref{ln:deadline1} and \ref{ln:deadline2}, which is invoked instead of the regular reaction code if and when the deadline is violated.

In this example, the federated reactors,  \texttt{Vision} and  \texttt{Robot}, can be deployed on separate machines and reactions in different reactors can be executed by multiple threads in parallel.
More importantly, for the same given input values and timing, the order of execution of reactions is always deterministic.
This modular design of reactors, determinism, and flexibility in deployments make \lfshort suitable for describing time-sensitive applications.

\subsection{Specifying Timing Behavior} 
\lf makes a distinction between two timelines, \textbf{logical time} and \textbf{physical time}~\cite{LohstrohEtAl:23:LogicalTime}.
\textbf{Logical time} is represented by a tag (timestamp and microstep in superdense time~\cite{Cataldo:06:Tetric,Maler:92:Hybrid}) that tracks the processing of events within the system.
It is a marker for the sequence and timing of events as understood by the system's logic.
\textbf{Physical time} tracks the actual movement of physical clocks, not to be confused with the conceptual clocks used in synchronous-reactive models.
In \lfshort's model of time, logical time, by default, lags behind physical time, meaning that the system's logical processing waits for the physical time to advance before proceeding.

\lf programs can explicitly specify a variety of timing behaviors in time-sensitive systems.
These timing behaviors include time-triggered events, communication delays, computation time, and deadlines.
The timing behavior specification in \lf allows deterministic execution for timed events and user inputs.
Reactions to events are executed in order based on the logical time (i.e., the tag).

The \textit{timer} in the \texttt{Camera} reactor (line \ref{ln:timer} in \figurename~\ref{fig:adas}) triggers periodic events with a specified offset and period.
The \textit{physical action} in the \texttt{EmergencyStop} reactor (line \ref{ln:physical}), represented in the diagram by a triangle with a ``P'', captures external asynchronous inputs, assigning them a logical timestamp based on the physical time, as measured by a local clock.
Lines \ref{ln:pedal_deadline} and \ref{ln:assist_deadline} give deadlines, which specify a maximum acceptable gap between the logical time and physical time of a reaction invocation.
These are also shown in the diagram as red markers in the reactions of the \texttt{Arm} reactor.

The \textbf{after delay} on line \ref{ln:after} is perhaps the most interesting timing specification in this program.
It increments the timestamp of the event conveyed along the connection from \texttt{Vision} to \texttt{Robot}.
This manipulation is used to ameliorate that effects of the fundamental tradeoff between consistency and availability in a distributed
system~\cite{LeeEtAl:23:CAL_CPS}.
In this case, it enables high availability at the \texttt{Arm} reactor, required to meet the 3 ms deadline, as long as the latency introduced by the \texttt{PedestrianDetector} and the communication does not exceed 10 ms.
To handle cases where the latency does exceed 10 ms, in this application we would likely use the \textit{decentralized coordinator} in \lfshort~\cite{BateniEtAl:23:Risk}, which treats such excess latency as a fault and provides for application-specific fault handling code to be executed.
In combination with deadlines, this mechanism offers sophisticated specification of timing requirements together with fault handlers to be invoked when those requirements are not met.

In short, \lf offers deterministic behavior under clearly stated assumptions, mechanisms to detect when these assumptions are violated, and fault handlers so that applications can react appropriately to violation of the assumptions.
This determinism applies even with parallel and distributed execution of \lfshort programs,
bringing the key advantages of determinism~\cite{Lee:21:Determinism}: repeatability, consensus, predictability (sometimes), fault detection, simplicity, unsurprising behavior, and composability.
For applications that require (or benefit from) nondeterminism, \lfshort includes explicitly nondeterministic constructs that can be used.
This is a notable contrast with most other concurrent and distributed computing frameworks, which give you nondeterminism by default and leave it to the designer to build deterministic behavior when needed.

\section{Lingua Franca and IEC 61499} 

The IEC 61499 architecture, according to the main website (\url{https://iec61499.com}), ``represents a component solution for distributed industrial automation systems aiming at portability, reusability, interoperability, reconfiguration of distributed applications.''
Like Lingua Franca, IEC 61499 is a component architecture, focusing on the interactions between concurrent and distributed components through message passing.
The standard is viewed by many as complementary to IEC 61131, which focuses on deterministic, cycle-driven, periodic computation.
IEC 61499 is event driven, like \lfshort, making it more suitable for applications like the one in the previous section.
But unlike \lfshort, the semantics of IEC 61499 is ambiguous in that a program may have more than one correct behavior~\cite{Cengic:06:DistributedControl}.
A project that could be pursued in the future would be to create a well-defined interpretation of the IEC 61499 that uses the deterministic reactor semantics of LF.

\section{Application to distributed safety} 

One of the applications that came up early in the project was an architecture for a networked emergency stop functionality, like that sketched in Section \ref{sec:lf}.
Today, many pieces of machinery are required to have easily accessible functionality to execute an emergency stop.
See, for example, Siemens safety applications with the S7-1200 FC CPU~\cite{Siemens:22:Safety}, which details 24 scenarios for safety door and emergency stop applications and outlines how these achieve performance levels (PL) compliant with the ISO 13849-1 standard and safety integrity levels (SIL) that are part of the IEC 61508 standard (and its related industry-specific standards, such as IEC 62061 for machinery and IEC 61511 for process industries).
These standards require dedicated wiring with current loops so that faults in the wiring can be detected.
For many modern machines, however, there are two key problems with this approach.
The first is that machines work in concert, and the shutdown process may need to be orchestrated with other machines.
The second is that dedicated wiring for each emergency function creates additional cost and points of failure.

The idea that this project began to pursue was to determine how to achieve comparable levels PL/SIL compliance with only packet-switched network connections.
Modern techniques like network clock synchronization, the use of heartbeats, and new innovations in the theory of distributed systems suggest promising approaches.
This effort was not directly pursued in this project because current safety standards and techniques (e.g. PROFIsafe) are well established in the industry, and their corresponding implementations are bound to certification and therefore managed conservatively. 
However, it remains an area of interest for the Berkeley team, who have pursued an effort based on a related concept developed by engineers at ABB~\cite{Johansson:20:Heartbeat}.
Followup work from ABB~\cite{Johansson:23:CAL} is based on a theoretical framework developed by the Berkeley team~\cite{LeeEtAl:23:CAL_CPS} and has led to collaboration with Swedish researchers that are pursuing formally verified implementations of a safety protocol.
Hence, we believe there is great potential for follow-up work in this area.

\section{Modes in Lingua Franca -- Behavior Trees} 

\lf supports \textit{modal models}~\cite{Schulz-Rosengarten:23:Modal}, where a reactor has modes of operation, and switching between modes is governed by a state machine.  There are alternative ways of describing decision logic in programs.

Behavior Trees (BTs) provide a lean set of control flow elements
that are easily composable in a modular tree structure. They are
well established for modeling the high-level behavior of non-player
characters in computer games and recently gained popularity in
other areas such as industrial automation.
While BTs nicely express control, data handling aspects so far
must be provided separately, e. g. in the form of blackboards. This
may hamper reusability and can be a source of nondeterminism.
In \cite{schulzrosengarten2024behaviortreesdataflowcoordinating} the authors propose a dataflow extension to BTs that explicitly models data relations and communication, and implement and validate that
approach in \lf.

The proposal of augmenting BTs with dataflow is, to the authors' knowledge,
the first attempt to do so systematically at the level of a coordination language. The aim is to combine the best of two worlds that,
so far, have seen little interaction through the involved research
communities or in actual practice. The authors argue that these concepts can
be of mutual benefit. Compared to ordinary BTs, the approach improves modularity and ensures determinism by replacing rather unstructured blackboards with a clean dataflow notation. Conversely,
dataflow formalisms can harness the intuitive, compact BT machinery that by now is proven in practice in a large and still growing
community of users in game development, robotics control, industrial automation, etc. With \lf as the basis for a concrete realization of this proposal, the authors leverage its deterministic semantics for concurrent, distributed real-time systems. Moreover, \lf’s polyglot nature makes the proposal compatible with a wide range of target languages.

\section{Converging complex and logical computation: Industrial automation use case in manufacturing} 

\begin{figure}
    \centering
    \includegraphics[width=1\linewidth]{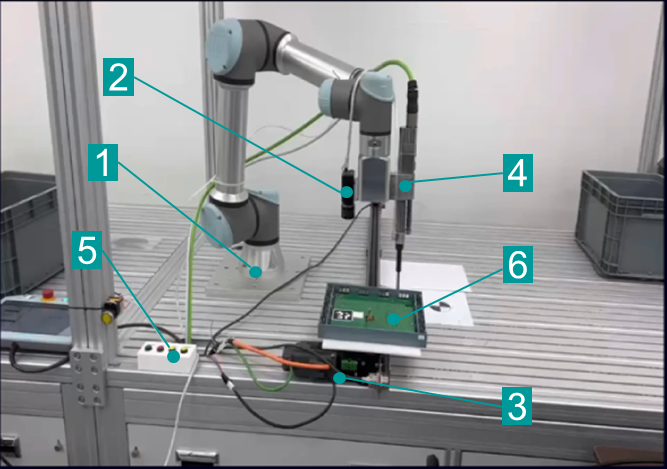}
    \caption{Hardware Setup}
    \label{fig:hardware}
\end{figure}

In some industry application scenarios, the computation is not only focused on simple logic calculation, but also involved with complex computation, for example, vision computation and robot arm control computation. Based on those potential requirements, we have designed a use case which involves a PLC, camera, robot arm, and additional components. The hardware is shown in Figure \ref{fig:hardware}.
It has several components:
\begin{enumerate}
    \item UR robot;
    \item 2D camera;
    \item PLC based conveyor system;
    \item Screw fastening machine;
    \item I/O button for screwing; and
    \item HMI board requiring a screw.
\end{enumerate}
The use case procedure is described in the following:

\begin{itemize}
    \item The Conveyor System conveys the HMI Panel back and forth to mimic a dynamic proceeding process.
    \item The eye-in-hand 2D Camera dynamically detects the target screw hole on the HMI board.
    \item An LF-based control system receives the data from 2D Camera, then based on algorithm of the PID controller, generates the speed of target robot.
    \item The UR robot moves along with HMI Board, arriving at a suitable position and then using the I/O button control Screw Fastening machine to insert the screw.
\end{itemize}

The control system has been designed by using LF. Its graphic top-down design makes the control system easy to analyze and track. The system contains several modules. The system diagram is  shown in Figure 2 and composes the following reactors:

\begin{figure}
    \centering
    \includegraphics[width=1\linewidth]{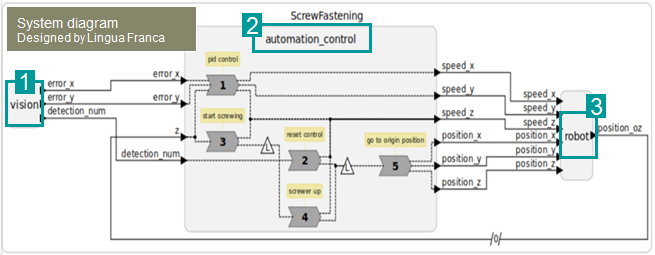}
    \caption{System Diagram}
    \label{fig:system}
\end{figure}

\begin{itemize}
    \item The \texttt{vision} reactor sends out the deviation (\texttt{error\_x},\texttt{error\_y}) of the target hole and sends out \texttt{detection\_num} to indicate whether the target hole is found. These three values will be send to reactor \texttt{automation\_control}.
    \item The reactor \texttt{automation\_control} receives the data from the \texttt{vision} reactor as well as feedback from the \texttt{robot} reactor. It contains the PID algorithm and motion adjustment algorithm. This module will send out robot’s speed to the \texttt{robot} reactor.
    \item The \texttt{robot} reactor receives the data from \texttt{automation\_control}. It will tell the robot how to move. It also sends out the \texttt{z} axis measurement of the robot to \texttt{automation\_control} to judge when the screw has been inserted.
\end{itemize}

\section{Concept for Princeton Future of Automation Lab} 

The Princeton Future of Automation Lab is one of the ``living labs'' of Siemens Technology.  It is a miniature factory with robots and automation systems to validate, demonstrate, and benchmark various automation technologies. It has various heterogeneous devices including: 

\begin{itemize}
    \item Dual arm robots built from two Kuka LBR iiwa robot arms;
    \item Magnemotion conveyor system;
    \item Universal Robots workstation;
    \item Gantry system based on SINAMICS; and
    \item various PLC, IPC, Edge, HMI devices.
\end{itemize}
The devices are connected mainly via fieldbuses (e.g. PROFINET). In most fieldbus technologies, there are different time synchronization technologies built in, such as PROFINET IRT, EtherCAT Distributed Clock, and EtherNet/IP CIP Sync. There is an ongoing effort to abstract fieldbus with middleware technologies to transition the lab into a more modern software defined automation system. How to unify the timing model across different devices with different fieldbus technologies will be a challenge.

\begin{figure}
    \centering
    \includegraphics[width=1\linewidth]{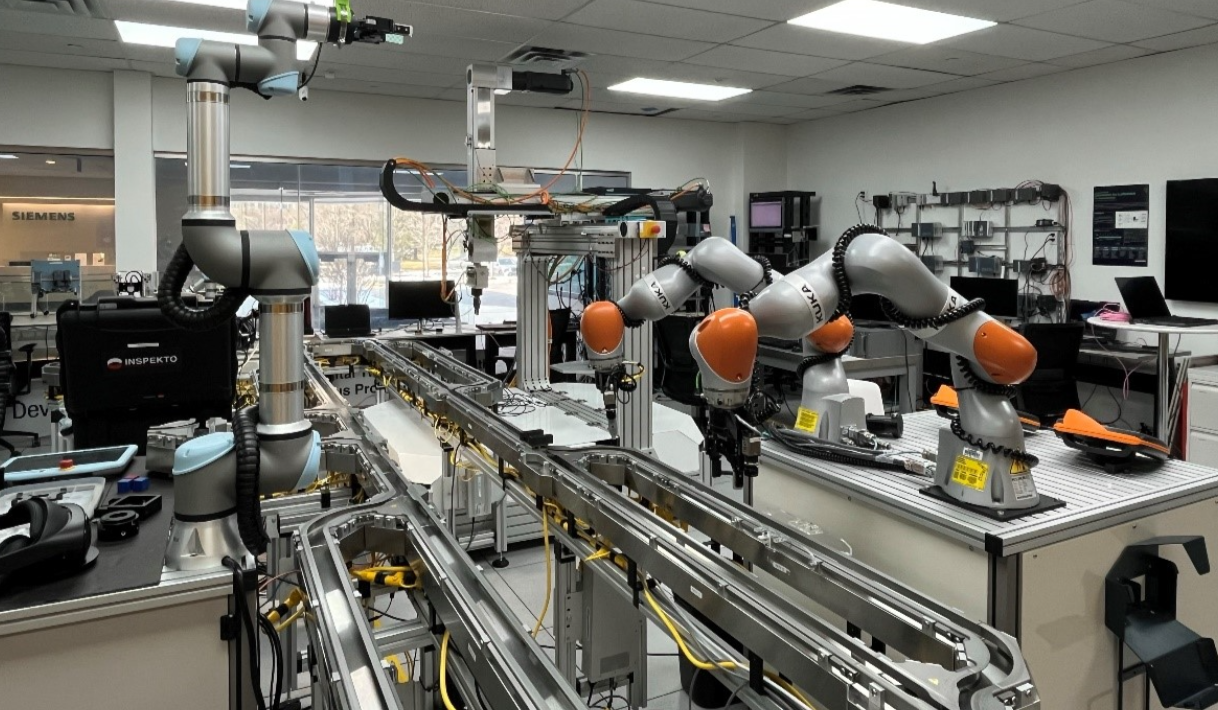}
    \caption{FoA Lab Overview}
    \label{fig:lab}
\end{figure}

In this regard, a \lf based coordination system could play a major engineering and coordination role, particularly when synchronization and timing are needed in the software defined automation system. Therefore, we are looking into use cases that can help to validate \lfshort for this purpose, concretely:
\begin{itemize}
    \item Conveyor tracking: the Magnemotion moves an object on the conveyor at variable speed and the robot arm tracks the object and attempts to grab the object while it is moving. This will involve synchronizing two different systems: Magnemotion and robot arm, together with the real time robot planning and movement control (e.g. PID loops). 
    \item Dual arms sync: the dual-arm Kuka robot can be used to evaluate the time sync of two arms, e.g. one arm as the master, the other as follower.
    \item Motion control: the gantry system in the lab is a multi-axis motion control system, where all the axes are precisely synchronized by the hardware so that the end-effector can follow a predefined path at a precise time. Currently this is done by PROFINET IRT and a hardware PLC, where the PLC program cycle is precisely synchronized with the fieldbus cycle. The challenge is to replace the PLC with a software implementation and determine how can multiple algorithms be precisely synchronized with the PROFINET IRT device.  Conceivably, this could be done by replacing the platform support code in the \lf runtime system that is responsible for timing with code that uses the PROFINET IRT device.
   
\end{itemize}
We see the potential benefit of using \lf to coordinate a software based system for automation and control in our lab. We will continue this effort from the Siemens side as we continue our lab related work now and in the future.

\section{Additional efforts motivated from this work and current results}

\subsection{Multicore} 

From the very beginning, \lf has provided deterministic parallel execution of programs on multicore machines~\cite{LohstrohEtAl:21:Towards} with impressive performance~\cite{MenardEtAl:23:DeterministicTACO}.
Each of the target language runtime systems provides a number of ``worker'' threads that, by default, matches the number of cores available.
Each worker thread executes reactions opportunistically with constraints on ordering that preserve determinism while maximizing parallelism.
This mechanism relies on an underlying operating system, such as Linux, that will execute threads on multiple cores when possible.

More recently, the \lfshort team has demonstrated that a similar mechanism can be used without an operating system~\cite{Berkun:24:MS}.
The number of worker threads is set to exactly match the number of cores, and each core executes one worker.
No operating system or thread library is required, making it possible to create deterministic multicore ``bare metal'' deeply embedded applications.

\subsection{Realtime behavior without RTOS} 

The UCB team has developed a layered scheduling strategy 
for Lingua Franca for enhanced real-time
performance that builds upon any priority-based
operating system thread scheduler. The application designers
need to specify \textit{only} the application-specific
deadlines, and the Lingua Franca runtime 
automatically converts them into appropriate
priority values for the OS scheduler to obtain
earliest deadline first scheduling~\cite{PaladinoEtAl:24:Sched}.
This technique promises to enable a generic Linux system to operate with real-time performance comparable to an RTOS.

\subsection{Deterministic scheduling} 
Scheduling is key to delivering \lfshort's deterministic semantics.
To achieve determinism, \lfshort constructs a dependency graph for reactions, such that at a given tag, all reactions must be scheduled in an order satisfying the dependency constraints.
The dependency graph is a partial order, which only specifies constraints necessary for determinism and leaves room for parallelism when possible.
In the default \lfshort runtime, the scheduler assigns levels to reactions based on the dependency graph.
At a tag, reactions are scheduled level-by-level, and those assigned the same level can be dispatched by the scheduler to two available workers that execute in parallel.

The default scheduling mechanism in \lfshort is a highly performant one~\cite{MenardEtAl:23:DeterministicTACO}. Yet, for certain hard real-time applications that demand provable, hard real-time guarantees, the opportunistic strategy to maximize parallelism could get in the way of analyzability. To address this issue, \lfshort introduces quasi-static scheduling fasciliated by a virtual machine called PretVM~\cite{lin2024pretvmpredictableefficientvirtual}, which aims to establish a one-to-one mapping, at compile-time instead of runtime, between a worker and a list of reaction invocations.
This approach offers greater analyzability, sufficient to prove system-level timing properties~\cite{schoeberl_et_al:OASIcs.WCET.2024.4}, at the cost of flexibility and performance offered by the default runtime. But for applications that present stringent and critical timing requirements, the quasi-static scheduling approach can be quite useful.

\section{Future research directions} 

\subsection{Open questions}


\subsubsection{Deterministic behavior for regression testing }

    Compared with the pattern with cycle-driven, periodic computation, an event-driven pattern has a higher degree of freedom and consequently adds complexity for the system designer. The stability of system still needs to be evaluated by system designer, and the evaluation still takes a considerable amount of time. \lf takes a big step towards simplifying this process by ensuring determinism, which enables regression testing.  Test patterns of event stimulus trigger one known-good behavior, and hence regression tests can be designed that check programs to ensure they continue to match that one known-good behavior as the program evolves. Moreover, \lfshort programs clearly distinguish between periodic and sporadic actions and enable specification of constraints on sporadic actions, such minimum time intervals. These features hold promise for building analysis tools that will help the designer find  potentially incorrect behaviors. The theory and design of such tools is a promising research topic for the future. 

\subsubsection{Pub/sub with determinism}
    
    Publish and subscribe (pub/sub) architectures are widely used in industry for control applications. Examples include: OMG-DDS, MQTT, OPC-UA Pub/Sub. The pub/sub pattern has advantages such as scalability, loose coupling (separation of concerns), flexibility, and eventually allowing to build a data-centric platform. However, the pub/sub pattern has some inherent drawbacks; in particular, it is intrinsically nondeterministic, meaning that a given program with a given input has more than one possible behavior. The question is, how can we strengthen the pub/sub pattern with a systematic approach without losing its advantages?

\section{Conclusion} 

The convergence of IT and OT and the trend of digitalization in industrial automation will likely change the corresponding technological landscape in the coming years. Two main challenges resulting from this transformation are:
\begin{enumerate}
    \item the introduction of new functionality for accessing and using engineering and operational data from the system to improve the system's operating qualities, and 
    \item achieving the shift from hardware-implemented to software-defined while preserving traditional characteristics of OT systems such as reliability, efficiency and predictability. 
\end{enumerate}

For the first challenge, data stemming from engineering and operations will need to be linked to precise contextual information for its consumption by analysis, optimization and diagnostic applications. Precise timing information is an essential component of this contextual information. This report has highlighted proven approaches for producing and providing both logical and physical time information across distributed systems, which are applicable to future industrial automation systems.

For the second challenge, software-defined automation paradigms will need to incorporate additional mechanisms to ensure the required operational qualities of industrial systems in the absence of hardware-provided guarantees. This report has presented approaches that can achieve deterministic execution and precise timing in distributed software systems while relaxing the requirements for the underlying hardware and communication resources. We believe that these approaches will make their way into critical software-defined automation systems in the future.





\bibliographystyle{plain}
\bibliography{Refs}

\end{document}